\documentclass[namedreferences]{solarphysics}

\usepackage[hyperref,optionalrh]{spr-sola-addons} 
\usepackage{graphicx}        
\usepackage{color}           
\usepackage{breakurl}        

\chardef\us=`\_

\def\arcsec{$^{\prime\prime}$}

\begin{document}
\sloppy

\begin{article}
\begin{opening}

\title{Locating Hot Plasma in Small Flares using Spectroscopic Overlappogram Data from the Hinode {\it EUV} Imaging Spectrometer}

\author[addressref={aff1,aff2},email={e-mail.louise.harra@pmodwrc.ch}]{\inits{L.K.}\fnm{Louise}~\lnm{Harra}\orcid{0000\-0001\-9457\-6200}}
\author[addressref=aff3] {\inits{K.}\fnm{Sarah}~\lnm{Matthews}\orcid{0000-0001-9346-8179}}
\author[addressref=aff3] {\inits{K.}\fnm{David}~\lnm{Long}\orcid{0000-0003-3137-0277}}
\author[addressref={aff4,aff5}] {\inits{T.}\fnm{Takahiro}~\lnm{Hasegawa}}
\author[addressref=aff6] {\inits{K.}\fnm{Kyoung-Sun}~\lnm{Lee}\orcid{0000-0002-4329-9546}}
\author[addressref=aff7] {\inits{K.}\fnm{Katharine K.}~\lnm{Reeves}\orcid{0000-0002-6903-6832}}
\author[addressref=aff5] {\inits{K.}\fnm{Toshifumi}~\lnm{Shimizu}\orcid{0000-0003-4764-6856}}
\author[addressref=aff8] {\inits{K.}\fnm{Hirohisa}~\lnm{Hara}\orcid{0000-0001-5686-3081}}
\author[addressref=aff9] {\inits{K.}\fnm{Magnus}~\lnm{Woods}\orcid{0000-0003-1593-4837}}

\address[id=aff1]{PMOD/WRC, Dorfstrasse 33
CH-7260 Davos Dorf,  Switzerland}
\address[id=aff2]{ETH-Z{\"u}rich, H{\"o}nggerberg Campus, Z{\"u}rich, Switzerland}
\address[id=aff3]{UCL-Mullard Space Science Lab., Holmbury St Mary, Dorking, Surrey, RH5 6NT, UK}
\address[id=aff4]{Department of Earth and Planetary Science, Graduate School of Science, The University of Tokyo, 7-3-1, Hongo, Tokyo, 113-0033, Japan}
\address[id=aff5]{Institute of Space and Astronautical Science, Japan Aerospace Exploration Agency, Sagamihara, Kanagawa, 252-5210, Japan}

\address[id=aff6]{CSPAR, The University of Alabama in Huntsville (Huntsville, Alabama, United States}
\address[id=aff7]{Harvard-Smithsonian Center for Astrophysics, 60 Garden St. MS 58, Cambridge, MA 02138, USA}

\address[id=aff8]{National Astronomical Observatory of Japan, 2-21-1 Osawa, Mitaka, Tokyo, 181-8588, Japan}

\address[id=aff9]{Lockheed Martin Solar and Astrophysics Laboratory, Palo Alto, CA, USA}

\runningauthor{Harra et al.}
\runningtitle{Locating Hot Flare Plasma}

\begin{abstract}
One of the key processes associated with the ``standard" flare model is chromospheric evaporation, a process during which plasma heated to high temperatures by energy deposition at the flare footpoints is driven upwards into the corona. Despite several decades of study, a number of open questions remain, including the relationship between plasma produced during this process and observations of earlier ``superhot" plasma. The {\it Extreme ultraviolet Imaging Spectrometer} (EIS) onboard Hinode has a wide slot that is often used as a flare trigger in the He~{\sc ii} emission line band. Once the intensity passes a threshold level, the study will switch to one focussed on the flaring region. However, when the intensity is not high enough to reach the flare trigger threshold, these datasets are then available during the entire flare period and provide high-cadence spectroscopic observations over a large field of view. We make use of data from two such studies of a C4.7 flare and a C1.6 flare to probe the relationship between hot Fe~{\sc xxiv} plasma and plasmas observed by {\it Reuven Ramaty High Energy Solar Spectroscopic Imager} RHESSI and {\it X-ray Telescope} XRT to track where the emission comes from, and when it begins. The flare trigger slot data used in our analysis has one-minute cadence. Although the spatial and spectral information are merged in the wide-slot data, it is still possible to extract when the hot plasma appears, through the appearance of the Fe~{\textsc xxiv} spectral image. It is also possible to derive spectrally pure Fe~{\sc xxiv} light curves from the EIS data, and compare them with those derived from hard X-rays, enabling a full exploration of the evolution of hot emission. The Fe~{\sc xxiv} emission peaks just after the peak in the hard X-ray lightcurve; consistent with an origin in the evaporation of heated plasma following the transfer of energy to the lower atmosphere. A  peak was also found for the C4.7 flare in the RHESSI peak temperature, which occurred before the hard X-rays peaked. This suggests that the first peak in hot-plasma emission is likely to be directly related to the energy-release process. 

\end{abstract}
\end{opening}

\section{Introduction}
     \label{S-Introduction} 
Solar flares release energy of the order of 10$^{30}$ ergs in tens of minutes. The energy required to power the flare is built up in the magnetic field of the active region over time, and then the ``standard" flare model predicts that the energy is rapidly released via magnetic reconnection. The released energy heats the plasma very quickly to hot temperatures, and leads to the acceleration of particles both down to the lower atmosphere, where they can drive chromospheric evaporation, and out into the heliosphere. Two of the key indicators of whether this model is a good description of reality are the ratio of thermal to non-thermal energy and the timing of the appearance of these signatures. If the model is a good description then the non-thermal energy must be sufficient to generate the thermal emission, and non-thermal signatures should precede the peak of the thermal energy. While there remain many observational gaps that constrain our ability to completely determine both components, X-ray and EUV spectroscopy provide a useful means to probe the early phases of the flare in detail.

``Hot" ($>$ 15 MK) plasma has previously been spectrally analysed using the {\it Yohkoh Bragg Crystal Spectrometer} instrument, which provided excellent temporal resolution \citep{Culhane1994}. However, this was a full Sun instrument, so no spatial information was available. The Extreme ultraviolet Imaging Spectrometer \citep[EIS:][]{EIS} onboard the \emph{Hinode} spacecraft \citep{Hinode} and the \emph{Interface Region Imaging Spectrometer} \citep[IRIS:][]{IRIS} both provide spatially resolved spectra in hot lines such as Fe~{\sc xxi} and Fe~{\sc xxiii} \citep[see, e.g.,][]{Polito2016,Lee2017}. \citet{Polito2016} in particular showed that the hot emission is co-spatial with the flare ribbons at the peak of the chromospheric evaporation. However, they rastered the slit of \emph{Hinode}/EIS across the field of view in order to provide spatial information in their study, which takes $>$ five minutes. In this work we demonstrate that we can extract the spatial location of spectrally pure hot Fe~{\sc xxiv} emission with better temporal resolution by using the slot data. 

Active region NOAA~12297 was famous for being the source of the first ``super geomagnetic storm of Solar Cycle 24" \citep{Wu2016}. The flare and eruption that created this superstorm occurred on the 15~March~2015 at 02:10~{\sc UT}. In this article, however, we specifically concentrate on two smaller flares: a C4.7 flare on 10 March 2015 and a C1.6 flare on 13 March 2015. The location of the flaring in the two small flares that we study is spatially coincident with the region that brightens in the larger flares that follow. Similar behaviour of recurrent flares was discussed by \citet{Polito2017}, who analysed a series of C-flares leading to an M-flare,  which was eruptive. A continuous hot plasma structure was seen in these flares, and they argue that the flares were all generated by the recurrence of bald-patch reconnection. The study of smaller recurrent flares with similar morphology may thus provide important clues about how the magnetic configuration evolves to facilitate larger energy release. Spectroscopy is an important tool for diagnosing where energy release and deposition occur within the magnetic-field configuration, but it often sacrifices temporal for spatial resolution. This work reduces those limitations by focussing on analysing the Hinode/EIS wide-slot data (overlappogram), which are able to provide higher temporal cadence over a large field of view. \citet{Widing} illustrates the power of overlappogram data  by the NRL XUV spectroheliograph on Skylab. They were able to analyse shifts and brightenings in the hot Fe~{\sc xxiii}--Fe~{\sc xxiv} emission lines during a flare, which continued after the non-thermal emission. In this article we concentrate on the Fe~{\sc xxiv} emission.

The Hinode/EIS overlappogram data have not been extensively used for science purposes yet and are typically used as an onboard flare trigger.  The spectrometer mode changes to a flare study once a flare is detected. Consequently there is a wealth of observational data available in the early phases of solar flares. Whereas \citet{Harra2017} demonstrated that the wide slot data can be used to extract velocity information of the He~{\sc ii} emission line, here we concentrate on extracting information on the evolution of hot plasma from the Fe~{\sc xxiv} emission line in order to explore the evolution of hot plasma in two small recurrent flares. Both flares analysed are not intense enough to trigger the flare trigger, meaning there is full temporal coverage with the overlappogram data. These data, combined with Atmospheric Imaging Assembly (AIA) data from the \emph{Solar Dynamics Observatory} \citep[SDO:][]{Pesnell2012}  provide the location and temporal evolution of hot flare plasma with much higher cadence than is achievable with rastering. The datasets from the \emph{Reuven Ramaty High Energy Solar Spectroscopic Imager} (RHESSI) spacecraft \citep{RHESSI} provide the location of the nonthermal hard X-ray emission.  An independent estimation of the temperature of the flare hot plasma is determined through hard X-ray spectroscopy, appearance of the Fe~{\sc xxi} emission line in IRIS spectra and through filter ratios in the {\it X-ray telescope} (XRT) onboard Hinode \citep{XRT}.

\section{Analysis}

 \begin{figure}    
   \centerline{\includegraphics[width=0.8\textwidth,clip=]{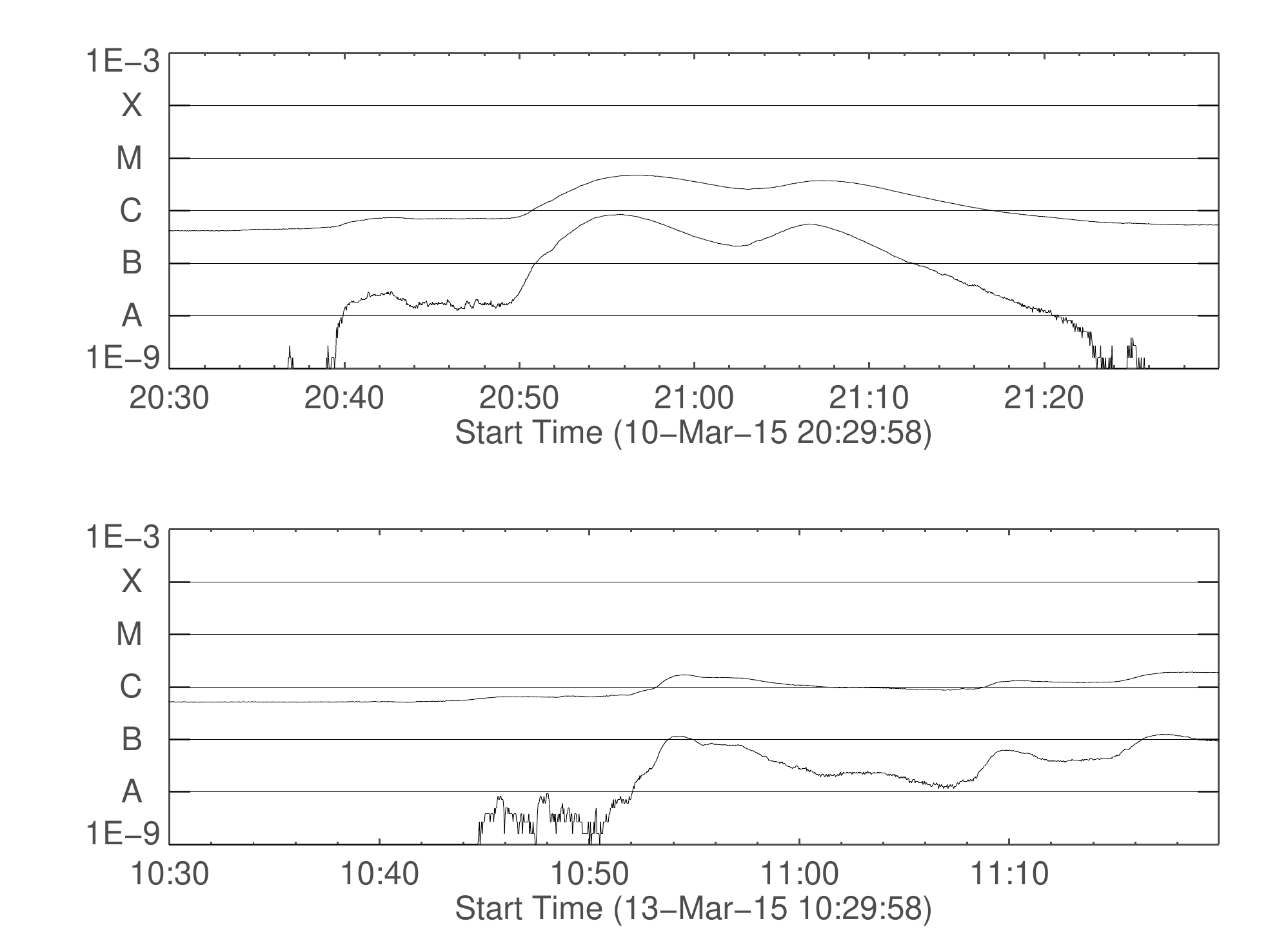}
              }
              \caption{ The GOES X-ray lightcurves for the two flares analysed in the 1-8 \AA\ passband (top) and 0.5-4.0 \AA\ (bottom) passband. }
   \label{goes}
   \end{figure}

\begin{figure}    
   \centerline{\includegraphics[width=0.8\textwidth,clip=]{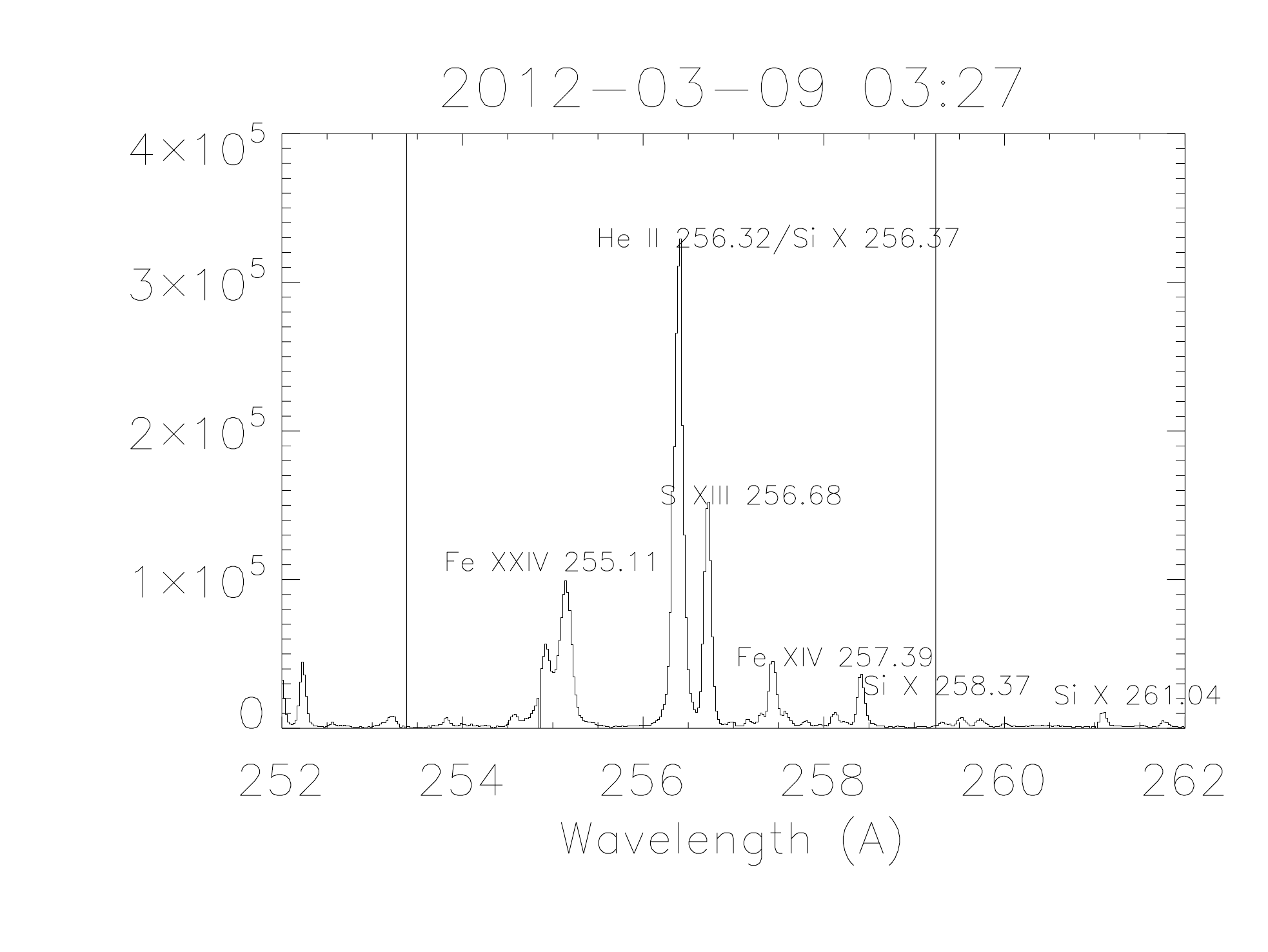}
              }
              \caption{ A slit spectrum during the rise phase of a flare in order to illustrate the emission lines. The vertical lines highlight the wavelength range captured in the wide slot data. From Harra et al., 2017 $\copyright$ AAS. Reproduced with permission.  }
   \label{spectrum}
   \end{figure}

Figure~\ref{goes} shows the GOES light curve for the two small flares studied (SOL2015\_03\_10T20:56 and SOL2015\_03\_13T10:54).  The Hinode/EIS instrument is an imaging spectrometer, which has two narrow slits that raster to build up images, and two slots. In this work we use the wide slot, which is 266{\arcsec} wide. The slot data provide spectral and imaging information over the entire active region at each exposure.  The EIS slot data are used as a flare trigger within the instrument and the mode changes to a slit raster once a large flare begins. The cadence of the EIS slot data is 60 seconds and the He~{\sc ii} emission line is used in order to capture early increases in intensity in flares. The He~{\sc ii} wide slot band includes a number of other emission lines that are generally weaker, but during flares can become prominent. These include Fe~{\sc xxiv}, S~{\sc xiii}, Fe~{\sc xiv} and Si~{\sc x}. Figure~\ref{spectrum} shows a slit spectrum illustrating what emission lines appear during the rise phase of a flare, as described in \citet{Harra2017}. When emission from these lines increases, they will appear as  additional images shifted in wavelength relative to the He~{\sc ii} image. The overlapping nature of the images is why the name ``overlappograms" was coined.  The dispersion direction is in the $x$ spatial direction. 

 In this work we demonstrate that we can extract the spatial location of spectrally pure hot Fe~{\sc xxiv} emission with better temporal resolution than the slit raster data by using the slot data. We compare the regions where we find the Fe~{\sc xxiv} to the Fe~{\sc xxi} spectra from IRIS.  IRIS was carrying out a flare watch study which consisted of a large, coarse four-step raster. The field of view was 6{\arcsec} $\times$ 119{\arcsec} with each raster taking 34 seconds. The field of view of the EIS slot data is 264{\arcsec} $\times$ 304{\arcsec}, enabling us to study the spatial and temporal evolution of hot plasma at multiple points in the flare region. The temperature of maximum abundance of Fe~{\sc xxiv} is 15\,MK, and for Fe~{\sc xxi} is 11\,MK. 
 
 In addition, RHESSI data \citep{RHESSI} are available for the both flares, allowing us to reconstruct the spatial distribution of the 6--12 and 12--25\,keV HXR emission using the CLEAN algorithm for detectors 1,3, 5, 7, 8 and 9 at two arcsec resolution. XRT data were also available for the smaller flare on 13 March 2015.  XRT data are calibrated using the SSW routine {\sf xrt\_prep}, which performs the standard dark subtraction,  removes the CCD bias, and telescope vignetting \citep{Kobelski}.  The filters used for the filter ratio calculation are Thin-Be and Thick-Al.  We use the response functions for these filters as described by \citet{Narukage}, and we account for the contamination layer on the CCD.  We note that a light leak in one of the front entrance filters of XRT occurred before these observations were taken, on 9 May 2012.  This light leak does not affect the throughput of the thin-Be and the Thick-Al filters (http://solar.physics.montana.edu/takeda/xrt\_straylight/xrt\_sl\_summary.html).
 
For the purpose of calibration and co-alignment, the SolarSoftWare routine {\sf aia\_prep.pro} is applied to the AIA data. The other instruments are aligned through comparison with the AIA data. For example, the EIS slot data and IRIS are aligned with AIA 304\,\AA\ imaging data. The Solar Aspect System and Roll Angle System onboard RHESSI provides pointing information \citep{Fivian}, and it has been established that the RHESSI disk center is known with an accuracy better than 0.2 arcseconds \citep{Batt}. Co-alignment between RHESSI and AIA was achieved through feature matching of the RHESSI sources to the ribbons observed in the 1600\,\AA channel on AIA. Given the 2 arcsecond resolution of the RHESSI image reconstruction, we estimate that the worst case error in the co-alignment to AIA is 2 arc seconds. The RHESSI--AIA co-alignment then allowed alignment to EIS following its alignment to AIA. For alignment of the IRIS data, we used the IRIS  slit-jaw image (SJI) 1330\,\AA\ and AIA 1600\,\AA\ images. The AIA images are selected as the nearest observing time of the SJI images, and the pixel resolution of the IRIS images is degraded to the same as the resolution of AIA images. For calculating the offset  between two images for the alignment, the IDL solarsoft routine {\sf cross\_corr.pro} was used.
   
 \section{10 March 2015, C4.7 flare}
 
\begin{figure}    
   \centerline{\includegraphics[width=1.2\textwidth,clip=]{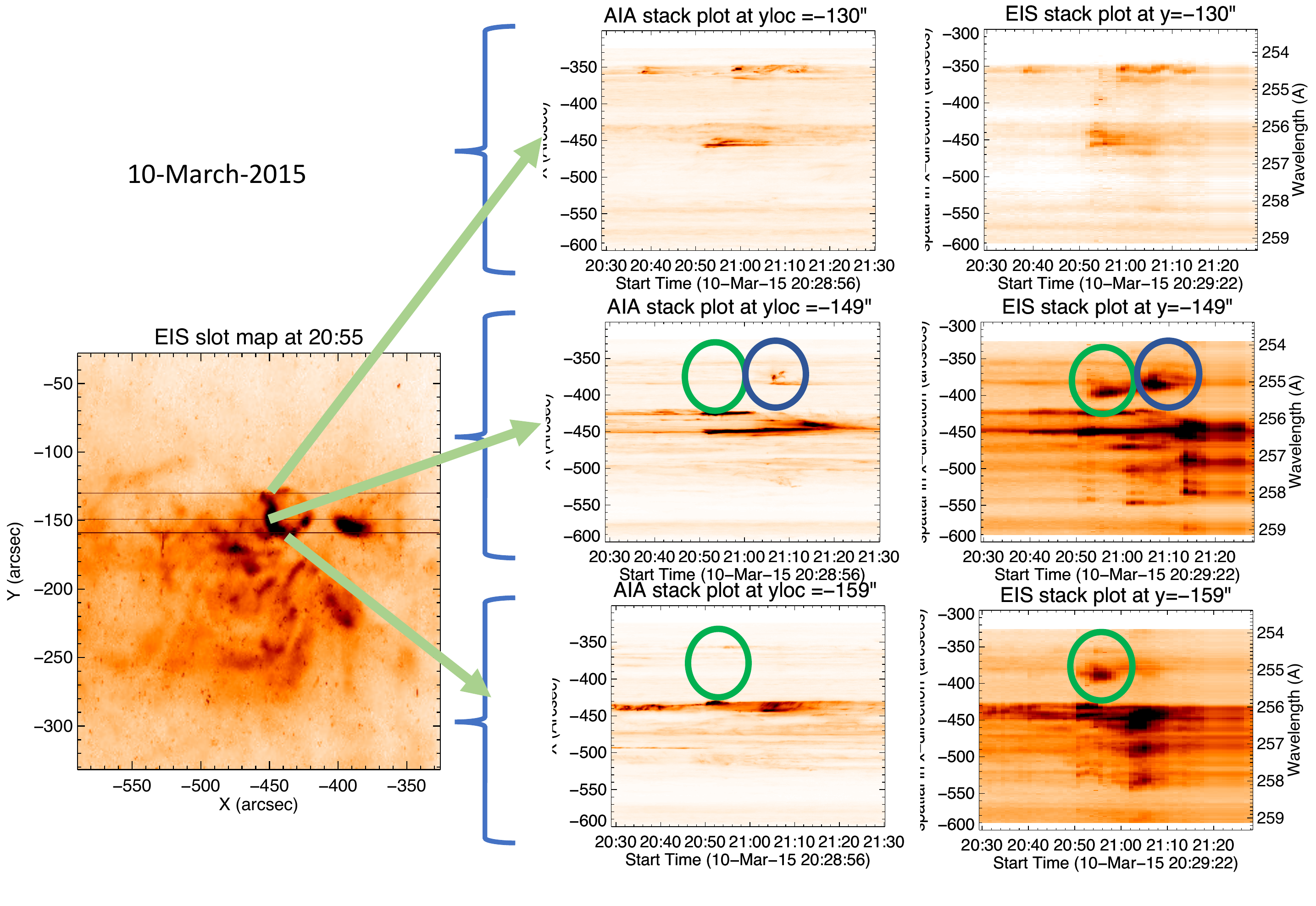}
              }
              \caption{ The left-hand image is a He~{\sc ii} slot image on 10 March 2015 at 20:55 UT which is during the first peak of the C4.6 flare. The stack plots are derived by choosing a position in the y-axes of the slot image, and plotting this position with time. Each stack plot has the position in the $x$-direction as the $y$-axes, with time on the $x$-axes. The left column is the AIA stack plot in the 304\,\AA\ filter at three positions that are highlighted on the slot image. The right column shows the EIS slot data at the same $y$-positions. The green circles highlight when the Fe~{\sc xxiv} spectral image appears in the EIS data, but is invisible in AIA confirming that it is a spectral feature. The blue circles highlight where emission is seen in both AIA and EIS, and hence the Fe~{\sc xxiv} is not found. The stack plots are shown in a reverse colour table where dark represents the highest intensity.  A movie is available in the Electronic Supplementary Materials accompanying this article (20150310\_304\_hires\_zoom.mp4) allowing the variability to be tracked in AIA, zoomed in to the flaring region. The EIS slot movie is available in the Electronic Supplementary Materials accompanying this article (10mar\_slotmovie.mp4). }
   \label{stack10}
   \end{figure}

 The X-ray lightcurve from GOES is shown in Figure~\ref{goes} {\bf top} panel. There is a small increase around 20:40 UT, and then the flare itself starts at 20:50 UT, with a peak at 20:56 UT followed by a second peak at 21:08 UT.   In order to determine whether additional ``features" in the slot images are spectral or spatial, we compare with the AIA He~{\sc ii} images. If the AIA images do not show a feature at the location of the slot images then this will be a spectral feature. Figure~\ref{stack10} shows a comparison of stack plots from AIA and EIS slot data at the same locations.  (Movies from both AIA and EIS are available in the Electronic Supplementary Materials accompanying this article).  Each stack plot shows one position in the $y$ direction on the slot image, plotted against time. The stack plots clearly show the regions that increase in intensity rapidly due to the flare. When comparing the AIA and EIS slot stack plots, it is clear that the EIS slot data are showing ``additional" images (this is also clearly seen in the movies). These are the spectral lines brightening that are shown in Figure~\ref{spectrum}. The wavelength scale is also shown to assist with interpretation.  The longest wavelength at the bottom of the scale and the shortest at the top with the He~{\sc ii} wavelength at 256.32\,\AA\ defined to be at the brightest part of the image.  The brightest part is seen in both AIA and EIS data. We also checked the 131\,\AA\ images and there was enhanced intensity at the same region. This emission highlights the ``hottest" emission in SDO/AIA focussed on the Fe~{\sc xvii} emission line. There are features in the EIS slot data that appear below this bright feature, and those are a blend of images from S~{\sc xiii}, Fe~{\sc xiv}, and Si~{\sc x} emission-line images. However, we are interested in the Fe~{\sc xxiv} emission,  and this lies above the He~{\sc ii} emission by approximately 50 pixels at 255.1\,\AA. The green circles highlight the times and positions when this emission is clearly seen in the EIS slot data and is missing in the AIA data. The AIA 304\,\AA\ band data include hotter emission lines in that band, but they will not be spectrally shifted. Hence if an additional image is seen in EIS but not in AIA, the  Fe~{\sc xxiv} emission location can be identified.  The location highlighted by the green circles shown in Figure~\ref{stack10} is the  Fe~{\sc xxiv} emission. The blue circles highlight where emission is seen in both AIA and EIS and hence no distinct Fe~{\sc xxiv} emission can be assumed. The slot image in Figure~\ref{stack10} shows the main bright image from which the green arrows emanate. At the $y$=150{\arcsec} and $x$=-380{\arcsec} position, there is an additional bright feature - this is the Fe~{\sc xxiv} image which is not seen in the AIA movie at 20:55 UT.  In the EIS slot movie, the size, intensity and evolution of the Fe~{\sc xxiv} feature can be tracked.

\begin{figure}    
   \centerline{\includegraphics[width=0.6\textwidth,clip=]{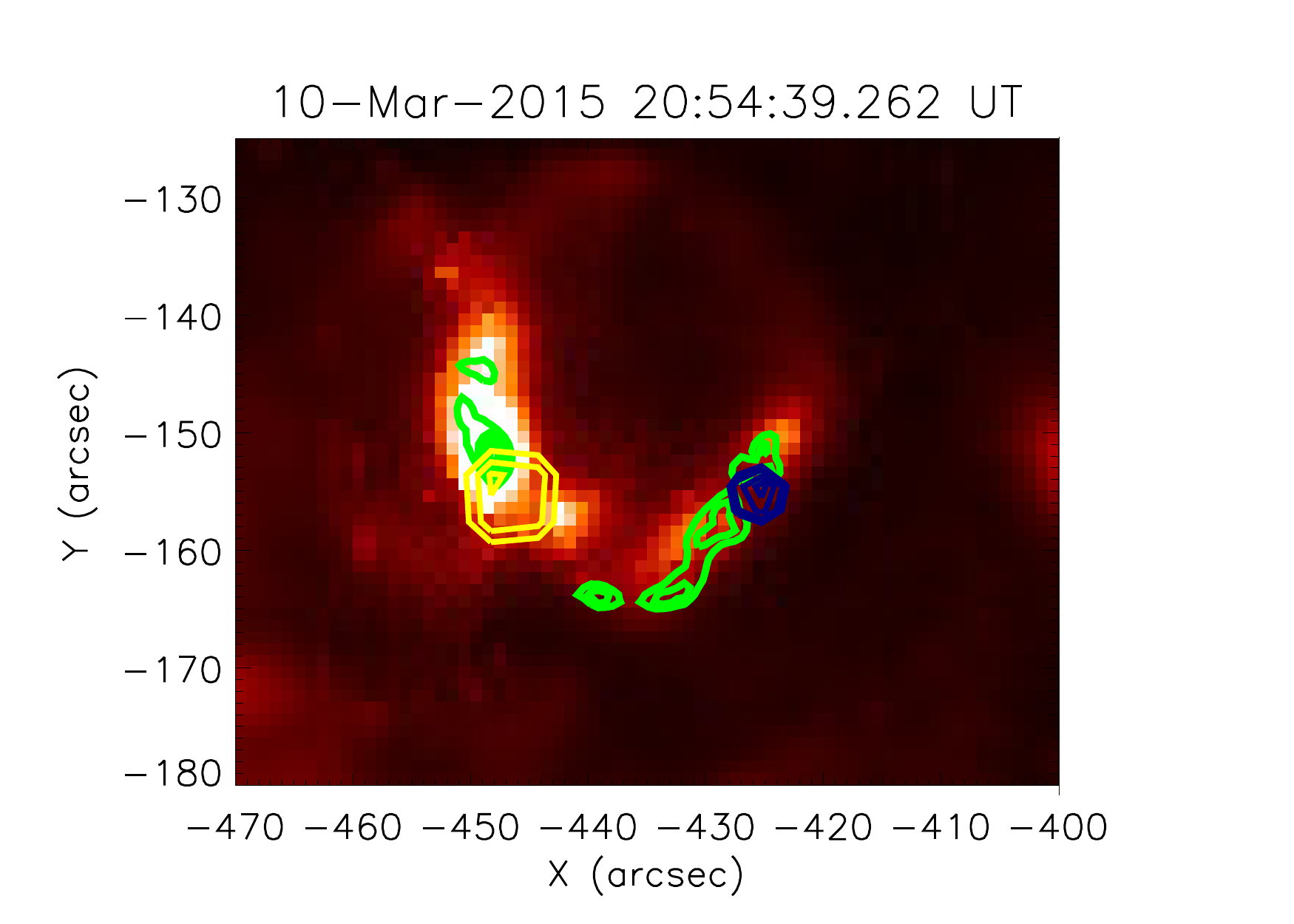}
              }
              \caption{EIS slot data (red) with AIA 1600\,\AA (green contours), the 12--25\,keV hard X-ray contours (yellow) and the 25--50\,keV hard X-ray contours are in blue for the 10 March 2015 C4.6 flare).  The Fe~{\sc xxiv} emission is located where the 12--25\,keV hard X-ray emission is.   }
   \label{hxr10}
   \end{figure}

 The location and timing of the hot plasma relative to the hard X-ray emission is central to understanding how well the ``standard" flare model describes the flare process. Figure~\ref{hxr10} shows the locations of the hard X-ray sources overlaid on the EIS slot data. The AIA 1600\,\AA\ emission is also shown as green contours to highlight the flare ribbons. The Fe~{\sc xxiv} emission is coincident with the 12--25 keV hard X-ray emission, shown as yellow contours.

\begin{figure}    
   \centerline{\includegraphics[width=0.9\textwidth,clip=]{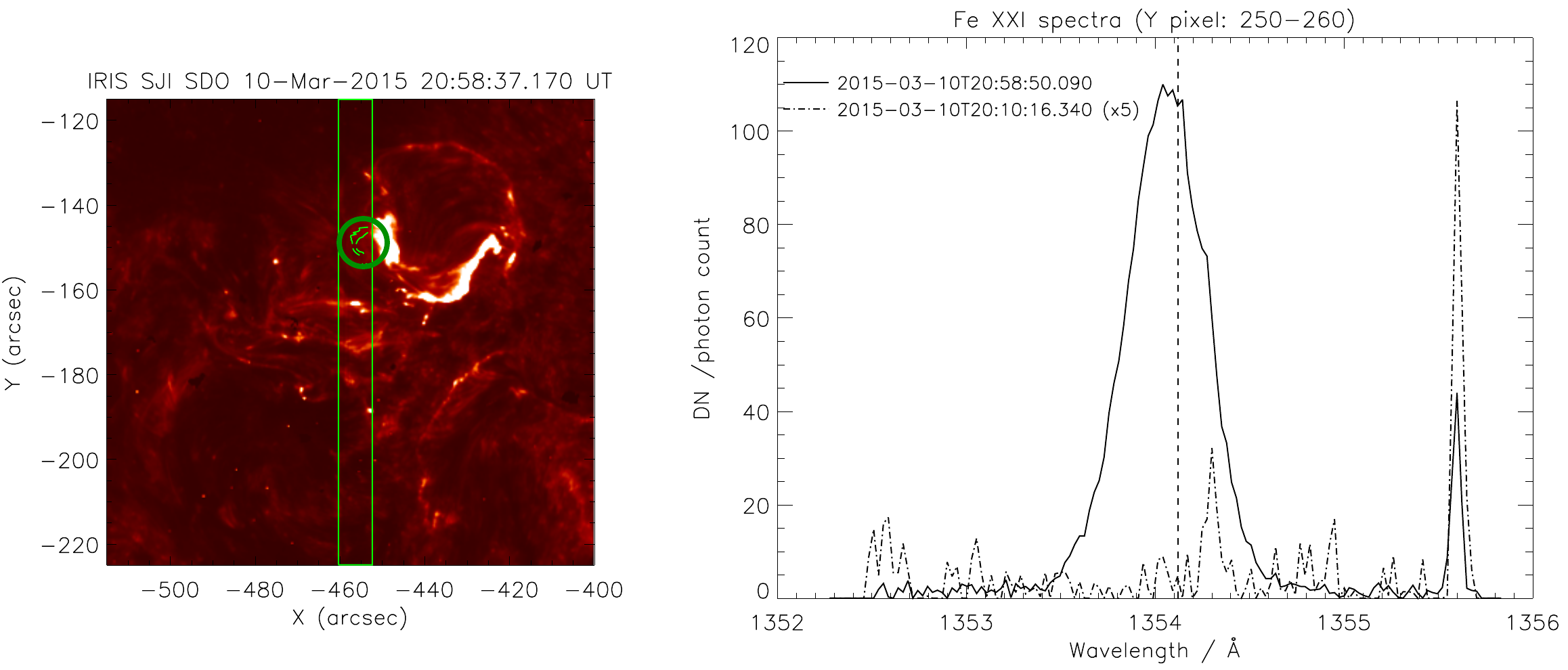}
              }
              
              \caption{Left: IRIS Fe~{\sc xxi} intensity contour and the raster FOV (green vertical lines) are overlaid on the SJI 1330. The green circle indicates the location of the Fe~{\sc xxi} emission.  Right: The averaged intensity of the circled location ($x$:2, $y$:250--260) before the flare (20:10:16 UT, dot--dashed line) and during the flare (20:58:50 UT, solid line). The vertical dashed line represents the reference wavelength of the Fe~{\sc xxi} 1354.12.    }
   \label{iris10}
   \end{figure}

 IRIS data were available during this flare, as is shown in Figure~\ref{iris10}. The location of the raster covered the edge of the flaring region, as is shown on the left-hand side of the figure. The right-hand side shows the Fe~{\sc xxi} spectral band before the flare and during the flare. The Fe~{\sc xxi} emission line appears during the flare. The IRIS  region slightly covers the left side of the Fe ~{\sc xxiv} region, which is showing the boundary part of the bright patch of Fe~{\sc xxiv} emission in EIS image. It confirms that the hot temperature plasma emission line appears during the flare at the same location that the EIS slot data is showing hot plasma. 

\begin{figure}    
   \centerline{\includegraphics[width=0.5\textwidth,clip=]{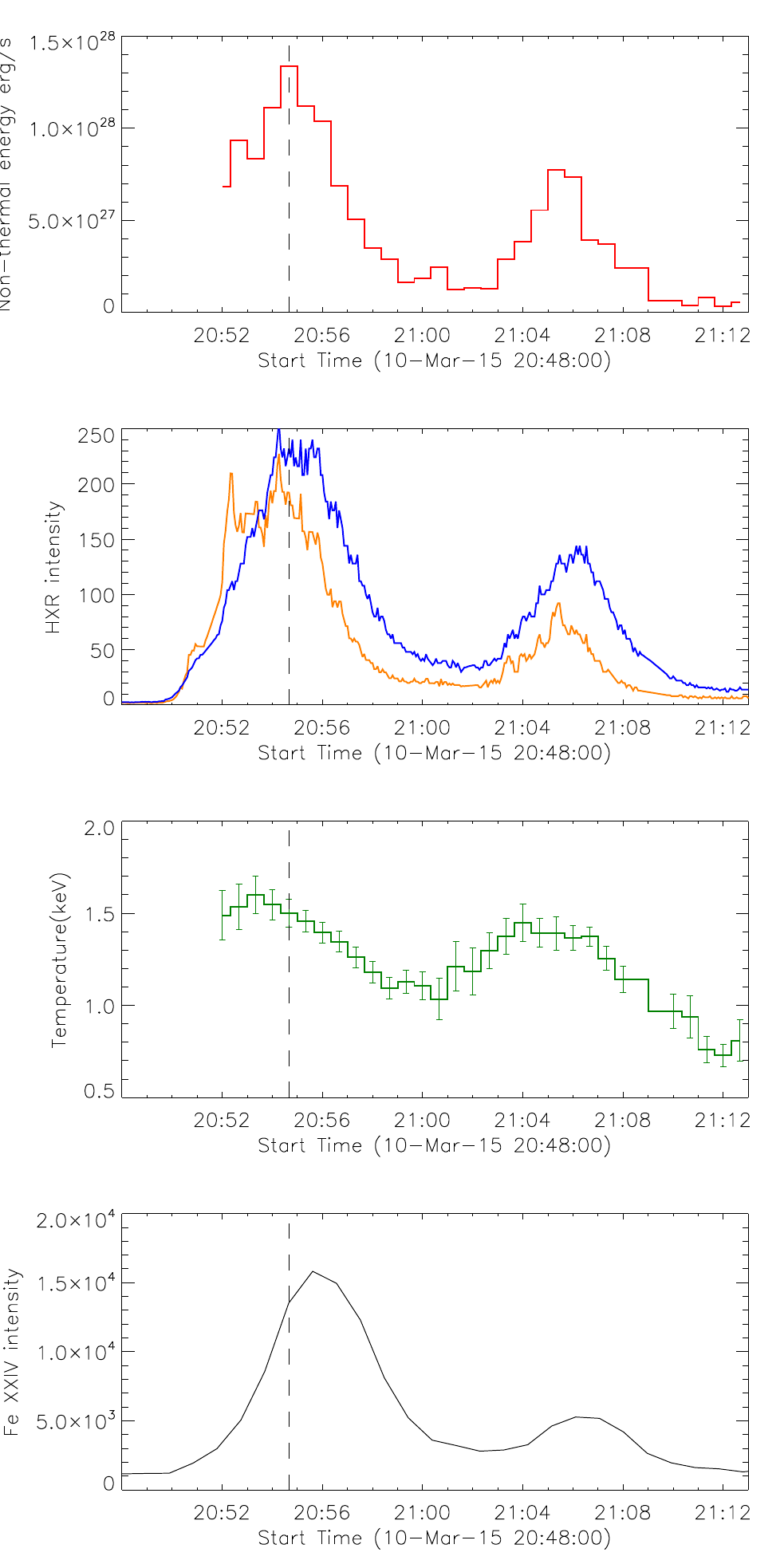}
              }
              \caption{ The non-thermal electron energy derived from RHESSI spectra (red); 6--12 keV intensity (blue) overlaid with 12--25 keV intensity (orange); plasma temperature (keV) derived from RHESSI spectra (green) and Fe~{\sc xxiv} intensity (black) from the EIS slot data for the C-flare. Plasma temperature peaks at 18.6\,MK at 20:53:20 UT, prior to the peak in non-thermal energy at 20:54:35 UT and the peak of the Fe XXIV intensity at 20:55:33 UT - indicating that the peak flare temperature occurs early on during the rise phase. A second energy release occurs in association with the C-flare, which shows similar timing behaviour, even though lower temperatures are reached }
   \label{10lcs}
   \end{figure}

In order to compare the Fe~{\sc xxiv} temporal behaviour with the hard X-ray data, the non-thermal energy and the temperatures were derived from the RHESSI data. These are shown in Figure~\ref{10lcs}. We plot overlays of the temporal variation of the HXR flux measured by RHESSI in the 6--12 keV (blue curve) and 12--25 keV (orange curve) energy ranges, together with the EIS slot Fe~{\sc xxiv} intensity (black line) and the non-thermal energy (red line) and plasma temperature (green) derived from the RHESSI spectra. The spectra from detectors 1,3,5,6,8 and 9 (with pulse pileup correction applied) were fitted at 40 second intervals using a combination of a thermal plus thick target spectrum. The non-thermal energy was calculated from the thick target fit parameters and peaks at 1.34$\times$10$^{28}$ erg s$^{-1}$, well correlated in time with the lower energy HXR flux i.e. 6--12 keV, and before the peak in Fe~{\sc xxiv} in both peaks of the flare.  The Fe~{\sc xxiv} emission lies on the flare ribbons.

 \section{13 March 2015, C1 flare}

\begin{figure}    
   \centerline{\includegraphics[width=1.2\textwidth,clip=]{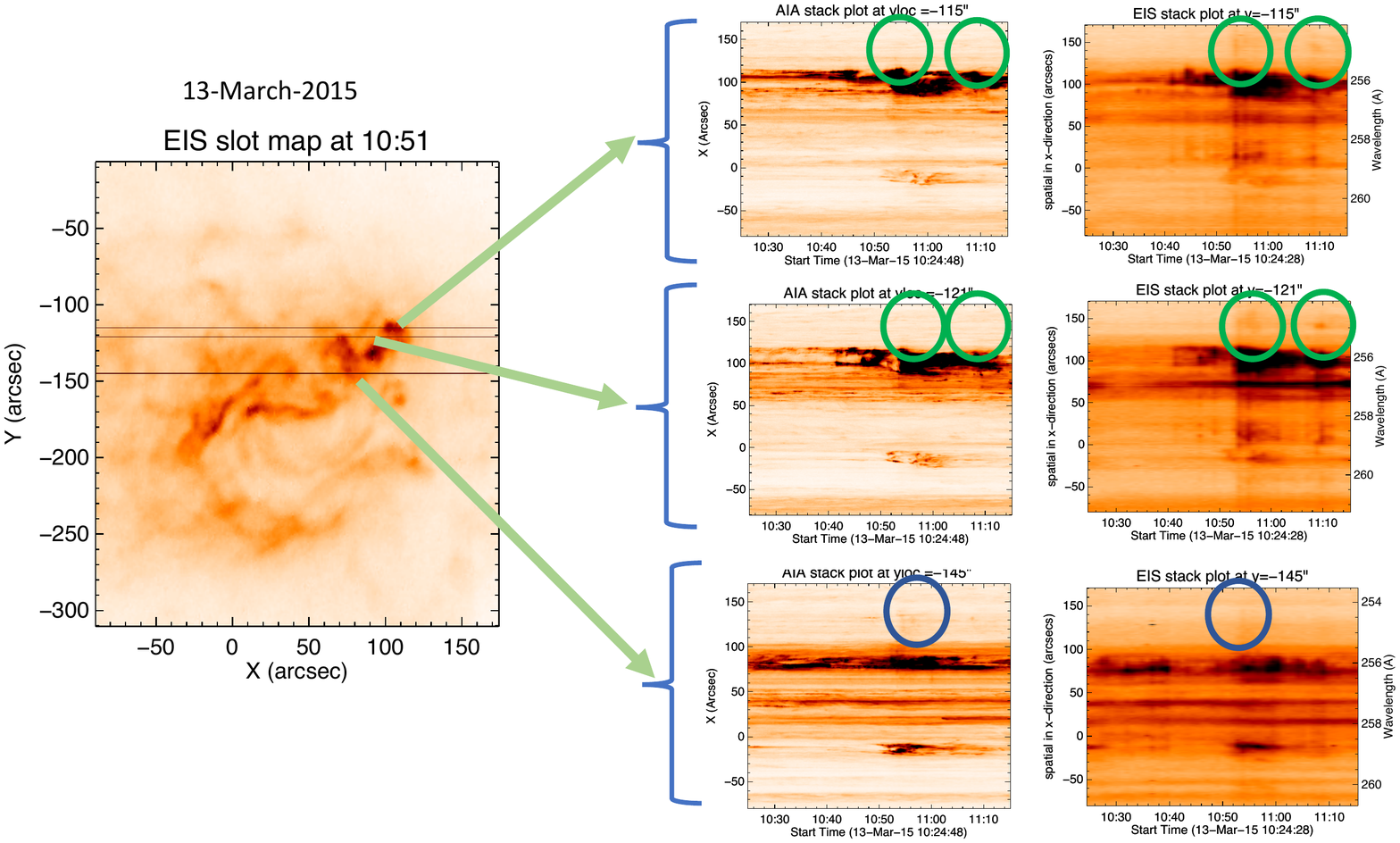}
              }
              \caption{ The left-hand image is a He~{\sc ii} slot image on 13 March 2015 at 10:51 which is close to the peak of the C1 flare. The stack plots are derived by choosing a position in the $y$-axes of the slot image, and plotting this position with time. Each stack plot has the position in the $x$-direction as the $y$-axes, with time in the $x$-axes. The left column is the AIA stack plot in the 304\,\AA\ filter at three positions that are highlighted on the slot image. The right column shows the EIS slot data at the same $y$-positions. The green circles highlight when the Fe~{\sc xxiv} spectral image appears in the EIS data, but is invisible in AIA confirming that it is a spectral feature. The blue circles highlight regions where there are features seen in both AIA and EIS and hence there is no distinct evidence for Fe~{\sc xxiv} emission. The stack plots are shown in a reverse colour table where dark is the highest intensity. A movie is available in the Electronic Supplementary Materials accompanying this article (\textsf{20150313\_304\_hires\_zoom.mp4}) where the variability can be tracked in AIA, zoomed in to the flaring region. The EIS slot movie is in the Electronic Supplementary Materials accompanying this article (\textsf{13mar\_slotmovie.mp4)}.   }
   \label{stack13}
   \end{figure}

\begin{figure}    
   \centerline{\includegraphics[width=0.8\textwidth,clip=]{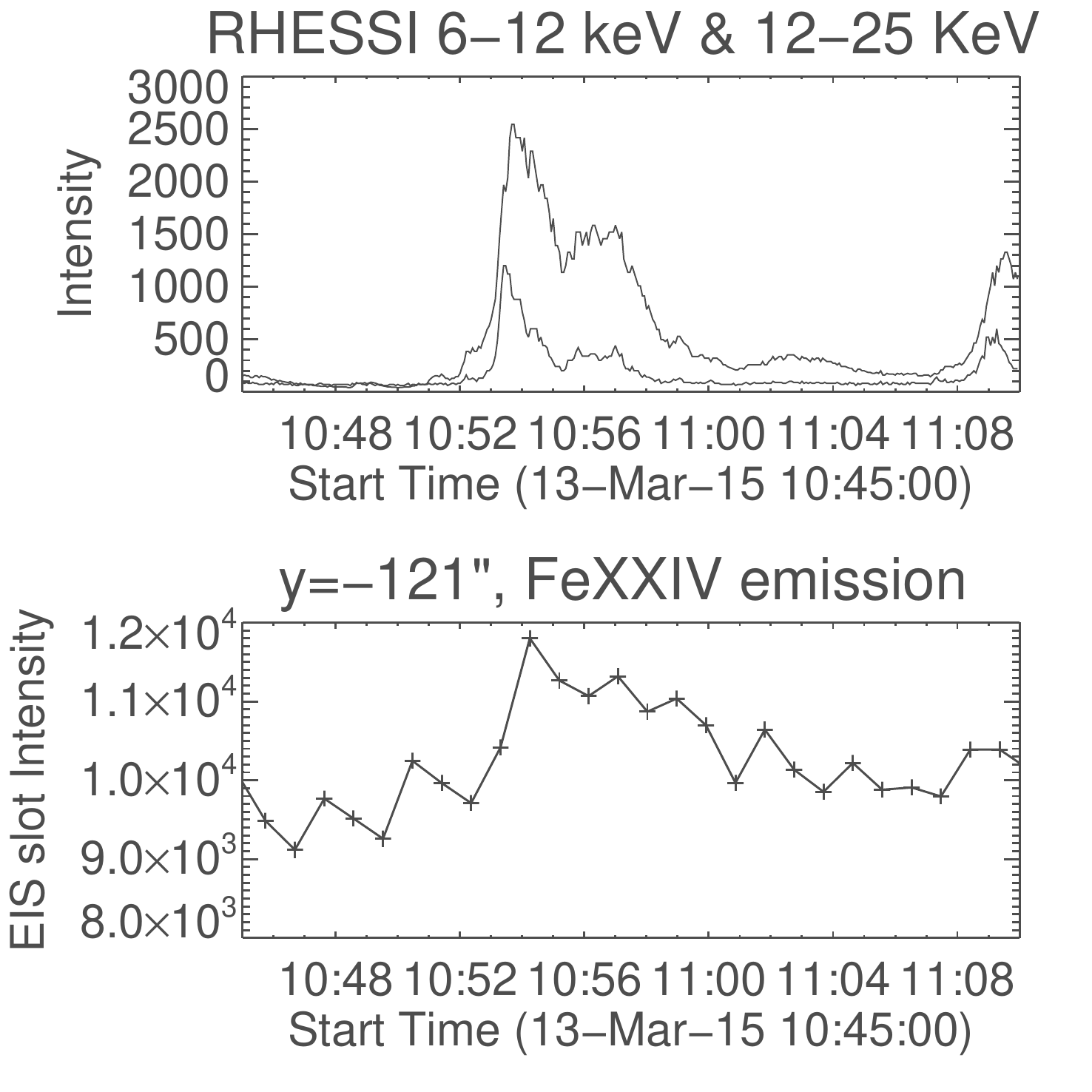}
              }
              \caption{ The top panel shows the 12--25\,keV HXR intensity, while the bottom figure shows the Fe~{\sc xxiv} intensity for the position $y$=121\,\arcsec stack plot. The Fe~{\sc xxiv} emission is weak but shows a clear enhancement when the flare begins.
 }
   \label{13lcs}
   \end{figure}

We study the 13 March 2015 C1 flare in order to determine whether hot emission can still be observed in such a small flare and whether its location and temporal evolution with respect to the nonthermal emission is similar to that observed in the previous flare. Figure~\ref{goes} (bottom panel) shows the lightcurve of the C1 flare on 13 March 2015.  Figure~\ref{stack13} shows the equivalent stack plots for the 13 March flare. It is noticeable that although there is distinct additional Fe~{\sc xxiv} emission seen in the EIS slot data, it is of lower intensity. Figure~\ref{13lcs} shows the Fe~{\sc xxiv} lightcurve along with the hard X-ray lightcurve.  From this figure it can be seen that, as in the previous flare, the Fe~{\sc xxiv} emission peaks after the hard X-ray emission observed in both the 6--12\,keV and 12--25\,keV channels. The intensity in the C1 flare is lower in both hard X-rays and the  Fe~{\sc xxiv} than for the 10 March 2015 C4.6 flare and we were unable to derive reliable spectral fits to the RHESSI spectra in order to determine the temperature from RHESSI. However, we know that temperatures of at least 15.8\,MK (peak of the G(T) for Fe~{\sc xxiv}) are seen in this small C1 flare. 
   
\begin{figure}    
   \centerline{\includegraphics[width=0.8\textwidth,clip=]{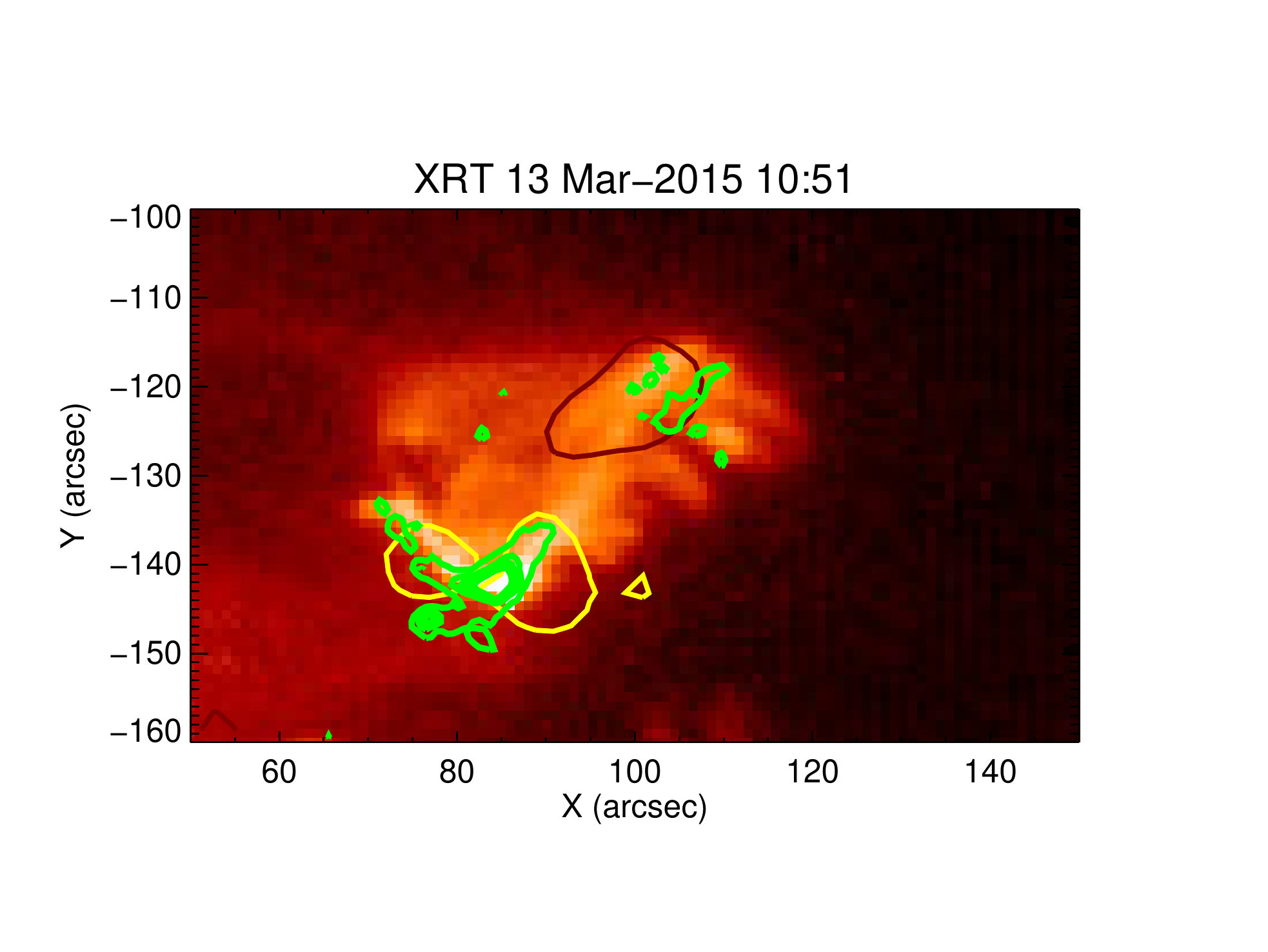}
              }
              \caption{ The XRT image at 10:51 UT with the RHESSI contours overlaid. The image is zoomed in to the main flaring region. The 6--12\,keV emission is shown as a dark red contour and the 12--25\,keV emission is shown as yellow.  The distinct Fe~{\sc xxiv}  EIS emission is seen at the 6--12\,keV hard X-ray source. The 12--25\,keV source has emission both in AIA and EIS at the Fe~{\sc xxiv}  location and hence pure Fe~{\sc xxiv}  emission cannot be extracted. The AIA 1600\,\AA\ data are highlighted as green contours. 
 }
   \label{13hxr}
   \end{figure}

Figure~\ref{13hxr} shows the XRT emission with the RHESSI 6--12\,keV and 12--25\,keV emission. The Fe{\sc xxiv} emission is $\approx$ 50\,\% weaker than the 10 March 2015 flare. The HXR intensity is also weaker by more than 50\,\%.  Although this is a small flare, it is clear that there is hot emission. It is associated with the north flare footpoint as highlighted by the 1600\,\AA\ emission. This was the location of the 6--12\,keV hard X-ray footpoint. Although there was a 12--25\,keV HXR source, there was no distinct  Fe{\sc xxiv} emission seen there as AIA was also seeing emission.  

\begin{figure}    
   \centerline{\includegraphics[width=0.8\textwidth,clip=]{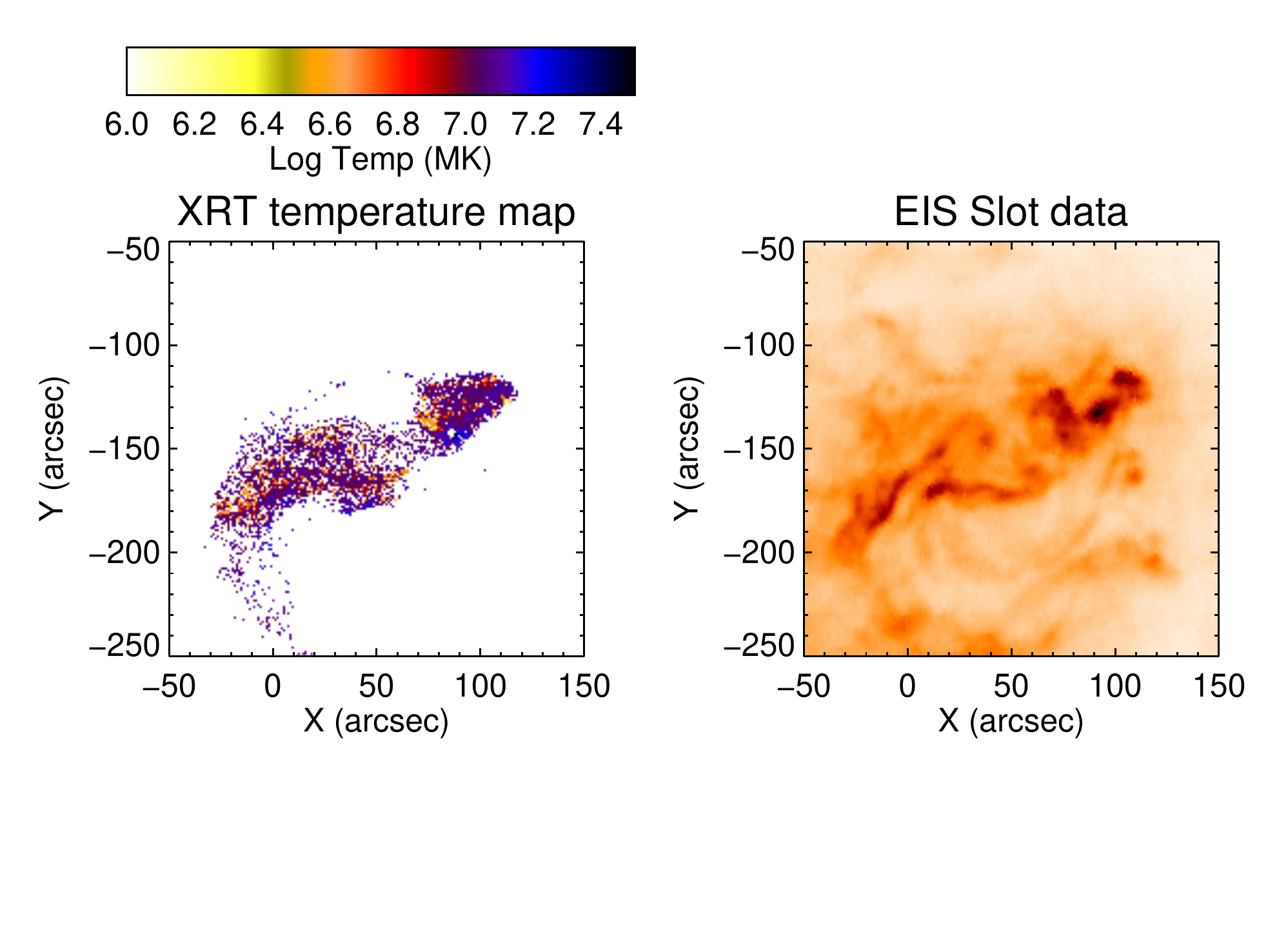}
              }
              \caption{The filter-ratio method was used with the XRT to create a temperature map. This is shown on the left-hand side. Pixels that had saturated data are shown as white.  On the right side is the EIS slot data. 
 }
   \label{xrt_eis}
   \end{figure}

\begin{figure}    
   \centerline{\includegraphics[width=0.9\textwidth,clip=]{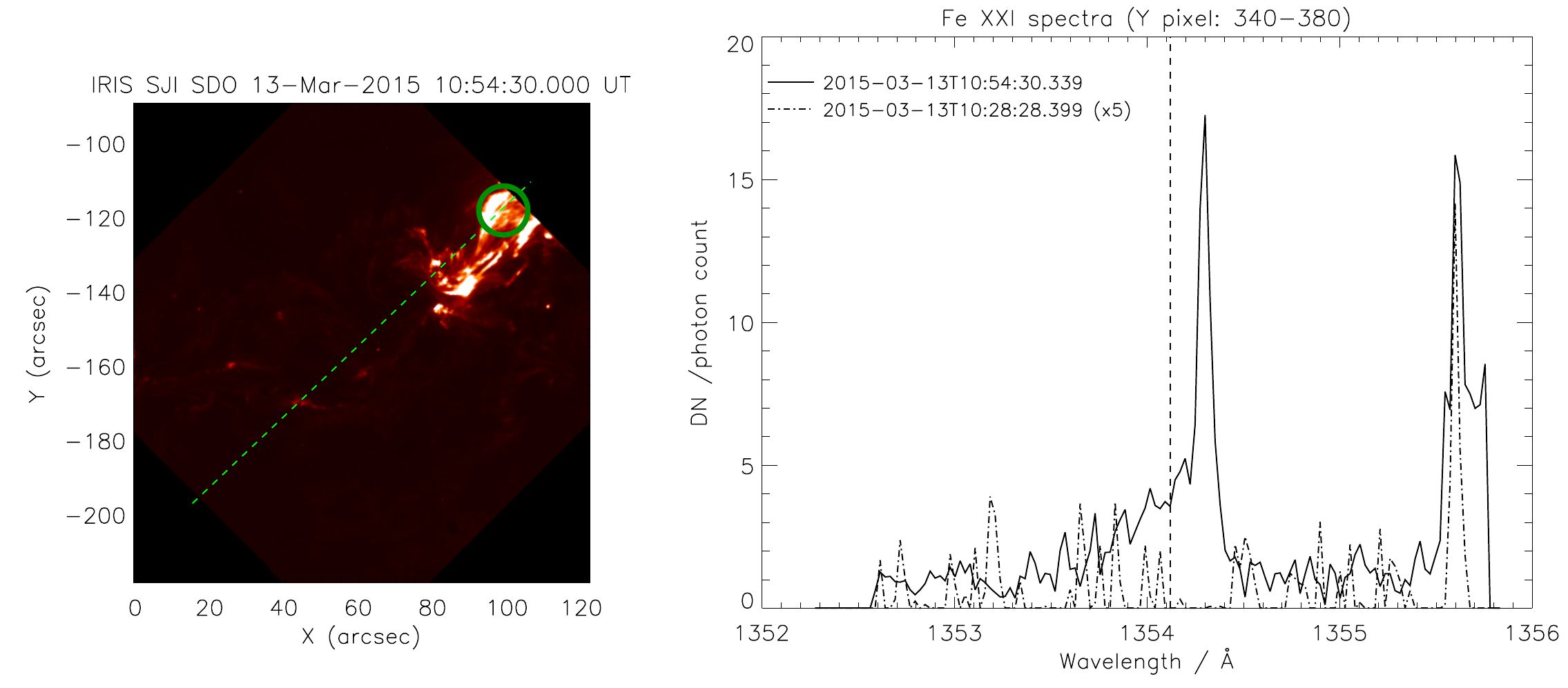}
              }
              
              \caption{Left: IRIS slit location (green dashed line) are overlaid on the SJI 1330. The position of the green circle is located the region where the Fe ~{\sc xxi} appears. Right: The averaged intensity of the circled location ($y$:340--380) before the flare (10:28:28 UT, dot--dashed line) and during the flare (10:54:30 UT, solid line). The vertical dashed line represents the reference wavelength of the Fe~{\sc xxi} 1354.12\,\AA line.       }
   \label{iris13}
   \end{figure}

 While the EIS slot data shows the likely existence of the Fe~{\sc xxiv} given the nature of the slot emission we wish to confirm this through other methods. In order to do this we perform a filter ratio analysis using the XRT data for the 13 March 2015 flare. We used the thin Be and thick Al filters, and assumed coronal abundances in Chianti version 8.0.0. The analysis showed that the hottest pixels are around the saturated area, which is the brightest part of the X-ray data shown in Figure~\ref{xrt_eis}.  There is no distinct Fe~{\sc xxiv} emission seen here due to the fact that there is also emission seen in AIA, and the Fe~{\sc xxiv} must be very weak if it is there. There is a restriction of this method. If the Fe~{\sc xxiv} emission appears in a location where there is other activity and it is weak then it is hard to observe. However, Fe~{\sc xxiv} emission is seen clearly in a different region, to the top right of the flaring region. The XRT filter ratio technique produce temperatures in the range from 10-20\,MK in this location. Complementary IRIS data also confirm the presence of hot plasma, and Figure~\ref{iris13} shows the location of the IRIS slit, and Fe~{\sc xxi} spectrum before and during the flare.  The emission is more than 5 times weaker in intensity than for the 10 March 2015 flare, but it is spatially coincident with the EIS Fe~{\sc xxiv} emission.

\section{Discussion}
The question of the spatial and temporal relationship between hot, thermal emission and nonthermal emission in flares is central to understanding the energy release mechanisms. Hot plasma is created during flares, reaching temperatures much higher than the ambient coronal temperature of a few MK. In the standard flare model it is generally accepted that hot plasma (between 5--25\,MK) is produced through the evaporation of  material following the bombardment of flare-accelerated electrons in the chromosphere (see a review by \citet{Fletcher2011}). In addition superhot  ($>$ 25\,MK) components have been observed, which may appear earlier in the flare. The {\it Yohkoh}/BCS instrument was capable of observing this superhot plasma through measurements of the Fe~{\sc xxvi} emission line indicating that temperatures of 30\,MK were reached \citep{Pike1996}. In their study, a range of GOES class flares showed Fe~{\sc xxvi} emission from C3.6 to M7.6. The smallest flare (C3.6)  in this sample showed a temperature of 22\,MK derived from Fe~{\sc xxvi}. \citet{Tanaka} derived temperatures using the Fe~{\sc xxvi}  and Fe~{\sc xxv} spectra from the Hinotori spacecraft during 13 large flares, and they found that the temperatures in the Fe~{\sc xxvi}  can increase to as much as 40\,MK.  

The highest temperature detected by RHESSI in the 10 March 2015 flare was 18.6\,MK, which occurred prior to the peak of the Fe~{\sc xxiv} emission, and prior to the peak in non-thermal energy, indicating that, as has been seen in previous studies, the peak flare temperature occurs early on during the rise phase. A second energy release occurs in association with the C-flare, which shows similar timing behaviour, even though lower temperatures are reached. While the timing relationship between the non-thermal energy peak and the EIS slot Fe~{\sc xxiv} peak is broadly consistent with the classic chromospheric evaporation scenario in which rapidly heated plasma is driven upwards following energy deposition by electrons, creating hot, dense plasma in the corona which radiates strongly, the early high temperature peaks observed by RHESSI suggest a different origin for this plasma, more directly related to the reconnection process and prior to the onset of explosive evaporation (e.g. \citealt{Caspi2014}; \citealt{Warmuth2016}).

\begin{figure}    
   \centerline{\includegraphics[width=0.9\textwidth,clip=]{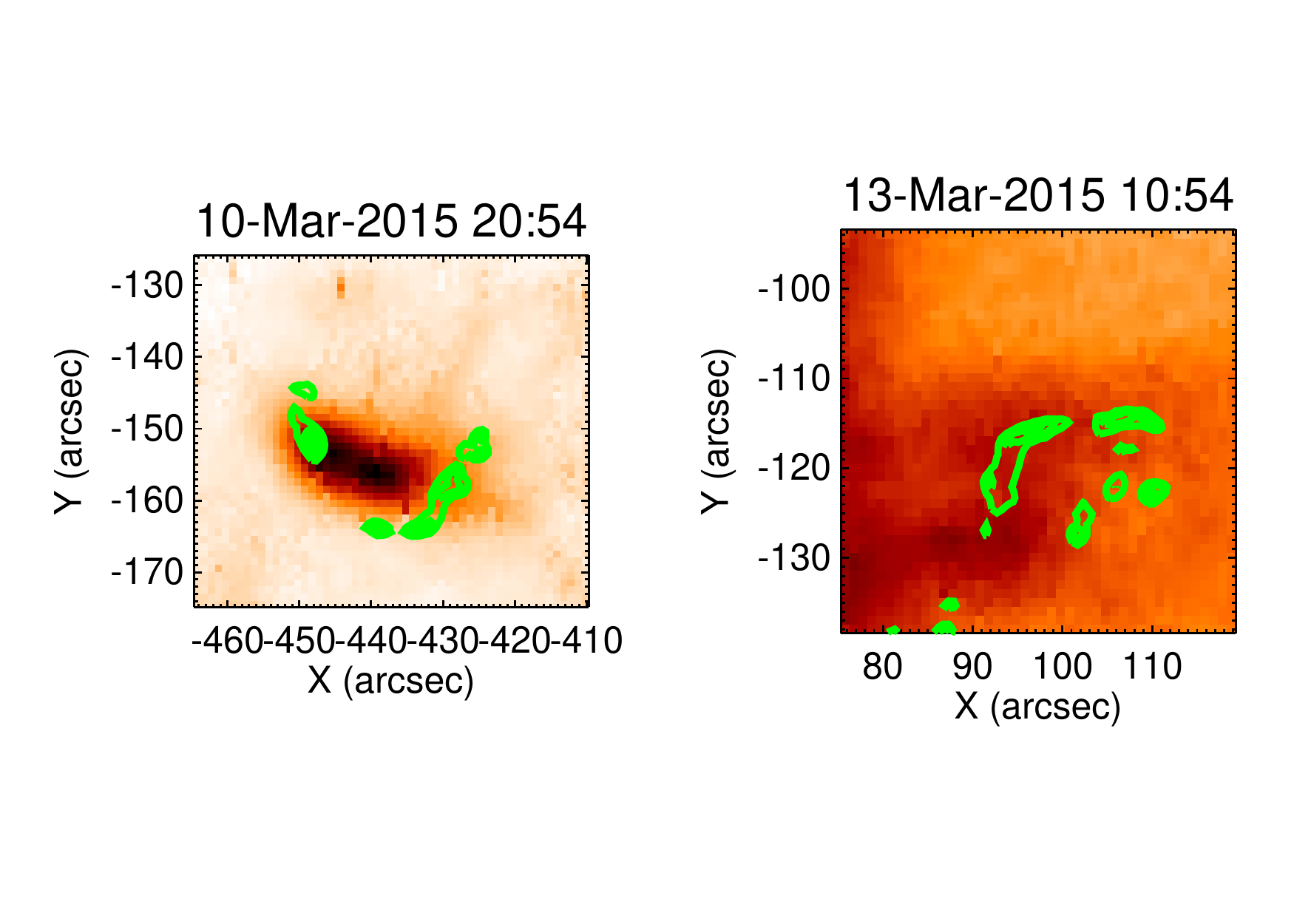}
              }
              
              \caption{Left:  The  Fe~{\sc xxiv} emission for the 10 March 2015 flare. This is   achieved by looking at the flaring region in the slot image, and then choosing the region of the image at the  Fe~{\sc xxiv} wavelength. Right:  the same pseudo Fe~{\sc xxiv} image for the 13 March 2015 flare. The emission at the  Fe~{\sc xxiv} location is weaker and also has some dynamical feature spatially overlapping with this region. However one can still see where the  Fe~{\sc xxiv} is located. Movies are associated with this figure are in the Electronic Supplementary Material accompanying this article (\textsf{fe24\_movie\_mar13.mp4} and \textsf{fe24\_movie\_mar10.mp4}.}
   \label{fe24_both}
   \end{figure}

The peak of the contribution function for Fe~{\sc xxiv} is 15\,MK, so it samples a similar temperature plasma to that detected by GOES, which is generally accepted to reflect the creation of hot, dense plasma in response to chromospheric heating. The EIS Fe~{\sc xxiv} observations allow us to demonstrate that this plasma also appears to be consistent with the onset of evaporation, while also letting us follow the changing morphology of the Fe~{\sc xxiv} emission by looking at the region in the slot image where the main image is shifted in wavelength from He~{\sc ii} at 256.32\,\AA\ to Fe~{\sc xxiv} at 255.11\,\AA. This is equivalent to a spatial shift of $\approx$ 50 pixels. 
Figure~\ref{fe24_both} shows a pseudo Fe~{\sc xxiv} image with the AIA 1600\,\AA\ image over-plotted (see also the accompanying online movies). The 10 March 2015 flare clearly shows strong  Fe~{\sc xxiv} emission. Interestingly, the  Fe~{\sc xxiv} emission is not static in one location but does move spatially with time. This behaviour has been observed for flare ribbons and HXR footpoints with complex dynamics often seen, but it is not widely reported for hot emission. The 13 March 2015  Fe~{\sc xxiv} emission is much weaker, and there is some contamination from other dynamical emission in that location, but a sense of the  Fe~{\sc xxiv} behaviour can also be extracted. In these two small flares studied in this article, we find that the hot emission is spatially coincident with the flare footpoints and very dynamic. Previous work by \citet{ryan} with Hinode/EIS has demonstrated that Fe~{\sc xiv} to Fe~{\sc xxiv} all show blue-shifts but with a significant stationary component that is inconsistent with the classic picture of evaporating plasma, while \citet{Young2015} demonstrate that observations from IRIS at higher spatial resolution show blue-shifted Fe~{\sc xxi} emission that occurs at and near the flare ribbon sites, consistent with the standard picture. In the flares studied here we see that the RHESSI observations for the 10 March 2015 flare indicate that there is an additional hotter plasma component which peaks in temperature prior to the peak of the non-thermal energy and Fe~{\sc xxiv} intensity, but that the hot Fe~{\sc xxiv} emission for both flares is located at the flare ribbons, consistent with the standard  evaporation scenario. This suggests that the Fe~{\sc xxiv} emission is a reliable indicator of the sites where chromospheric evaporation is occurring, and it likely represents the upper temperature limit of plasma produced by this process. Future work quantifying the dynamics of the spatially resolved Fe~{\sc xxiv} emission could provide valuable insights into the distribution of energy deposition sites as the flare progresses.

There have been suggestions over the years that superhot components with $>$30\,MK may be created through a different physical mechanism \citep{Caspi2010}. This is likely to be related to the acceleration region in the corona, since it occurs earlier in the flare process. 
In \citet{Caspi2014} they measure temperatures with both RHESSI and GOES for 37 strong flares to explore this.  They correlated the RHESSI derived temperatures with GOES classification and the same for maximum temperature derived from GOES with GOES class. Their study concentrated on M- and X-class flares. Only two of the M-flares achieved superhot temperature $>$ 30\,MK. If one extrapolates to the GOES classification for the two events in this article, one obtains a temperature derived from their RHESSI temperature $T_{\it R}$=9.8\,MK (for C1.6) and 16\,MK (for C4.7).  For the GOES temperature statistical study, a value of $T_{\it G}$=9.3\,MK (for C1.6) and 13\,MK (for C4.7) is produced.  The \citet{Caspi2014} study did not suggest that one could extrapolate to these small GOES classes, but it is interesting to compare. For the C1.6 flare, we do see Fe~{\sc xxiv}, which has the peak of the G(T) to be 15.8\,MK. This is similar to  that extrapolated from \cite{Caspi2014} using the RHESSI temperature measurements for the C4.7 flare.  The C1.6 measurements extrapolated from the RHESSI statistical study are lower than the peak of G(T) Fe~{\sc xxiv}. This is important when constraining models of flares and extrapolations to both small and larger stellar flares. The intensity of the EIS slot Fe~{\sc xxiv} emission was different in each flare, with the C4.7 flare peaking with a 50\,\% higher intensity than the C1.6 flare.  



The method described here, allows the existence of hot plasma in small flares, and its temporal evolution to be determined in a consistent way. There is a wealth of data available from the EIS wide slot studies, that is an excellent source for the ``hot plasma" detection. The temporal cadence here was one minute as this is data for the flare trigger. However, other EIS studies exist with cadences of ten seconds, and these can be used for further studies.

\begin{acks}
This project has been funded through the award of the Daiwa-Adrian Prize through the Daiwa Anglo-Japanese foundation. \emph{Hinode} is a Japanese mission developed and launched by ISAS/JAXA, collaborating with NAOJ as a domestic partner, and NASA and STFC (UK) as international partners. Scientific operation of \emph{Hinode} is performed by the \emph{Hinode} science team organized at ISAS/JAXA. This team mainly consists of scientists from institutes in the partner countries. Support for the post-launch operation is provided by JAXA and NAOJ (Japan), STFC (UK), NASA, ESA, and NSC (Norway). SDO data were obtained courtesy of NASA/SDO and the AIA and HMI science teams. DML acknowledges support from the European Commission's H2020 Programme under the following Grant Agreements: GREST (no.~653982) and Pre-EST (no.~739500) as well as support from the Leverhulme Trust for an Early-Career Fellowship (ECF-2014-792) and is grateful to the Science Technology and Facilities Council for the award of an Ernest Rutherford Fellowship (ST/R003246/1).
 \end{acks}

\section*{Disclosure of Potential Conflicts of Interest}
The authors declare that they have no conflicts of interest.


\bibliographystyle{spr-mp-sola}
\bibliography{sola_bibliography_hotplasma}  

\IfFileExists{\jobname.bbl}{} {\typeout{}
\typeout{****************************************************}
\typeout{****************************************************}
\typeout{** Please run "bibtex \jobname" to obtain} \typeout{**
the bibliography and then re-run LaTeX} \typeout{** twice to fix
the references !}
\typeout{****************************************************}
\typeout{****************************************************}
\typeout{}}

\

\end{article} 

\end{document}